\documentclass[fleqn,usenatbib]{mnras}

\usepackage{newtxtext,newtxmath}

\usepackage[T1]{fontenc}

\DeclareRobustCommand{\VAN}[3]{#2}
\let\VANthebibliography\thebibliography
\def\thebibliography{\DeclareRobustCommand{\VAN}[3]{##3}\VANthebibliography}

\usepackage{graphicx}	
\usepackage{amsmath}	

\title[Coma UDGs]{Probing ultra-diffuse galaxies out to the virial radius of the Coma cluster with \textit{XMM--Newton} }

\author[M. S. Mirakhor and S. A. Walker]{
M. S. Mirakhor\thanks{E-mail: msm0033@uah.edu}
and S. A. Walker
\\
Department of Physics and Astronomy, The University of Alabama in Huntsville, Huntsville, AL 35899, USA
}

\date{Accepted XXX. Received YYY; in original form ZZZ}

\pubyear{2015}

\begin{document}
\label{firstpage}
\pagerange{\pageref{firstpage}--\pageref{lastpage}}
\maketitle

\begin{abstract}
We probe the formation scenarios and AGN occupation fraction of ultra-diffuse galaxies (UDGs) in the nearby Coma cluster by utilizing \textit{XMM$-$Newton} observations of 779 out of 854 UDG candidates identified by Subaru survey. Their origin is probed by measuring the dark matter halo mass of the stacked sample of UDGs and the population of low-mass X-ray binaries residing in globular clusters. Our measurements suggest that the average UDG population does not have a substantial amount of hot gas or a large number of globular clusters. This supports the formation scenario, in which UDGs are puffed-up dwarf galaxies, agreeing with that obtained for 404 Coma cluster UDGs using \textit{Chandra}. We also determine the active galactic nuclei (AGN) occupation fraction of UDGs by cross-correlating the position of UDGs with the detected point sources in Coma. We detect three X-ray sources with detection significance $\sigma \geq 5$ that could be off-centre AGN within 5 arcsec from the centre of the UDG 317, UDG 432, and UDG 535. We identify an optical counterpart for the X-ray source associated with the UDG 317, suggesting that this source is more likely an off-centre AGN. Based on the current data, however, we cannot conclusively constrain whether the detected AGN is residing in the Coma cluster or not.

\end{abstract}

\begin{keywords}
galaxies: clusters: individual (Coma) -- X-rays: general -- X-rays: galaxies -- galaxies: formation 
\end{keywords}



\section{Introduction}
\label{sec: intro}
The presence of large, extremely faint systems in galaxy clusters is not a recent discovery \citep[e.g.][]{impey1988virgo}. However, deep imaging surveys of nearby galaxy clusters \citep[e.g.][]{mihos2015galaxies,munoz2015unveiling} have only recently revealed a copious population of faint systems that has remained relatively unexplored in shallower surveys. Perhaps the most interesting result is the discovery of a large population of massive but extremely faint galaxies in the Coma cluster \citep{van2015forty}, using the Dragonfly Telephoto Array \citep{abraham2014ultra}. These galaxies, which were dubbed as ultra-diffuse galaxies or UDGs, have very low central surface brightness ($\mu_{g,0}\gtrsim 24$ mag arcsec$^{-2}$) with large effective radii ($r_{\rm{eff}}>1.5$ kpc) similar to Milky Way-type galaxies.

While their existence is well established, it is still an open question of how UDGs fit into the general picture of galaxy formation and evolution. Despite the abundance of UDGs in galaxy clusters, their formation and evolutionary pathways can be constrained to two main formation scenarios \citep{kovacs2019constraining}. The first scenario is that UDGs are "failed" massive galaxies that were prevented from further star formation by gas removal \citep[e.g.][]{van2015forty,van2016high}. The second one is that they are genuine dwarf galaxies, and their large spatial extent is caused by strong feedback from supernovae or active galactic nuclei (AGN) activity \citep[e.g.][]{beasley2016overmassive,amorisco2016ultradiffuse,peng2016rich}. The former scenario suggests that UDGs reside in massive dark matter halos, while in the second scenario, UDGs are expected to reside in low-mass dark matter halos. This difference implies that one can constrain the formation scenarios of UDGs by measuring their dark matter halo mass.    

X-ray observations provide a powerful tool to probe the total mass distribution in galaxies. Several studies \citep[e.g.][]{babyk2018x,lakhchaura2018thermodynamic} found that the X-ray luminosity is proportional to the total gravitating mass of galaxies. Therefore, massive galaxies are expected to have higher X-ray luminosities than low-mass galaxies. \citet{kovacs2019constraining} employed \textit{XMM$-$Newton} observations to probe the formation scenarios for a sample of isolated UDGs by constraining their dark matter halos. No statistically significant emission from the individual UDGs or from the stacked sample of UDGs was detected. The absence of significant emission implies the lack of a massive dark matter halo. This, in turn, suggests that the bulk of isolated UDGs reside in low-mass dark matter halos.

More recently, \citet{kovacs2020ultra} probed the formation scenarios for a sample of UDGs residing in the nearby Coma cluster, using a similar analysis to that done in \citet{kovacs2019constraining}. Specifically, they carried out a \textit{Chandra} X-ray analysis on 404 out of 854 UDGs detected within the field of Coma by the Subaru Suprime-Cam survey \citep{yagi2016catalog}. The formation scenarios of UDGs were probed by measuring the X-ray luminosity of the hot diffuse gas around UDGs and low-mass X-ray binaries (LMXBs) residing in globular clusters (GCs). They also did not detect significant X-ray emission from the hot gas or from GC-LMXBs. Based on the upper limits on the X-ray luminosity, \citet{kovacs2020ultra} concluded that the bulk of the UDG population in the Coma cluster are genuine dwarf galaxies. This conclusion resonates well with that derived for isolated UDGs \citep{kovacs2019constraining}. 

Moreover, \citet{kovacs2020ultra} constrained the occupation fraction of AGN in UDGs by cross-correlating the position of UDGs with the identified X-ray sources in the Coma cluster. They only identified two UDGs that have off-centre source, which might be AGN, at distances of 3.0 and 3.2 arcsec from the centre of the UDG 317 and UDG 412. This implies an AGN occupation fraction of about 0.5 per cent, which is comparable to that found for low-mass galaxies. 

However, the reported results in the work of \citet{kovacs2020ultra} are based on 404 UDGs, which only represents around 47 per cent of the full list of UDG candidates detected in the Coma cluster. In this work, taking advantage of our high-quality \textit{XMM-Newton} data of the Coma cluster that cover the cluster out to the virial radius \citep{mirakhor2020complete}, we rely on a larger sample of UDGs to probe their formation scenarios and AGN occupation fraction. As our data cover a larger region of Coma, we have \textit{XMM$-$Newton} observations for 779 out of 854 UDG candidates in the Coma cluster, excluding the central 4 arcmin region. This number is almost twice the number of UDGs (404 UDGs) studied in \citet{kovacs2020ultra}, and represents around 91 per cent of UDG candidates identified in the Coma cluster. This large sample of UDGs would provide a better constraint on their origin and AGN occupation fraction.

The X-ray analysis adopted in the current work is essentially the same as the one by \citet{kovacs2020ultra}, but we now apply this analysis on 779 UDGs instead of 404 UDGs. As in \citet{kovacs2020ultra}, we also rely on the publicly-available data for 854 UDG candidates in the Coma cluster \citep{yagi2016catalog}. The spatial distribution of these UDGs cover a region of 1.7 $\times$ 2.7 deg$^2$ of the Coma cluster. As spectroscopic measurements are not available for a large fraction of them, we cannot conclusively determine whether these galaxies are members of the Coma cluster. However, based on the radial velocity measurements of sub-samples of UDG candidates, several studies \citep[e.g.][]{van2016high,van2017extensive} found that the vast majority of UDG candidates reside in the Coma cluster. We, therefore, consider that all UDG candidates reside in the Coma cluster, as is done in \citet{kovacs2020ultra}. 

Throughout this work, we adopt a $\Lambda$ cold dark matter cosmology with $\Omega_{\rm{m}}=0.3$, $\Omega_{\rm{\Lambda}}=0.7$, and $H_0=100\,h_{100}$ km s$^{-1}$ Mpc$^{-1}$ with $h_{100}=0.7$. At the redshift of Coma, 1 arcsec corresponds to 0.48 kpc. Uncertainties are at the 68 per cent confidence level, unless otherwise stated.

\section{Observations and data reduction}
In this work, we used 61 \textit{XMM$-$Newton} observations in the field of the Coma cluster that cover the cluster out to the virial radius with nearly complete azimuthal coverage. These observations were taken in the period between 2000 and 2019, and the total observation duration is around 2.3 Ms. The details of all \textit{XMM$-$Newton} observations used in this work are summarized in Table \ref{tab: xmm_observations}.

We reduced the X-ray data using \textit{XMM$-$Newton} Science Analysis System (XMM-SAS) version 18.0 and Current Calibration Files (CCF), following the methods described in the Extended Source Analysis Software (ESAS) cookbook\footnote{https://heasarc.gsfc.nasa.gov/docs/xmm/esas/cookbook/xmm-esas.html}, as is also done in \citet{mirakhor2020complete,mirakhor2020high}. We initiated the data processing by running the \textit{epchain} and \textit{emchain} scripts, followed by the ESAS tasks \textit{mos-filter} and \textit{pn-filter} to filter the data for soft-proton flares and produce event files for MOS and PN detectors. We screened the data in the MOS detectors for CCDs in anomalous states, and any affected CCDs were then excluded from further analysis. The ESAS source-detection tool \textit{cheese} was used to detect point sources and extended substructures. The next step in data processing is to create spectra, RMFs, ARFs, event images, and exposure maps for the entire region of interest using the \textit{mos-spectra} and \textit{pn-spectra} tasks. Then, by running the \textit{mos-back} and \textit{pn-back} tasks, the intermediate files created by \textit{mos-spectra} and \textit{pn-spectra} were turned into the quiescent particle background spectra and images in detector coordinates. 

We further examined the data for residual soft-proton contamination that may have remained after light-curve filtering. After the spectral parameters for the soft-proton contamination have been derived, the \textit{proton} task was run to produce images of the soft proton contamination in detector coordinates. We also applied an additional screening step by running the \textit{Chandra} tool \textit{wavdetect} to detect point sources and extended substructures within the field of view that were missed using the ESAS tool \textit{cheese}.   

After weighting the MOS and PN detectors for all of the \textit{XMM$-$Newton} observations by their relative effective area, the main components for a background-subtracted and exposure-corrected image produced from the analysis procedure described above were then merged and adaptively smoothed into a single image. Following \citet{kovacs2020ultra}, we created X-ray images in the soft (0.5$-$1.2 keV) and broad (0.5$-$7.0 keV) bands to account for the X-ray emission from hot gaseous halos and LMXBs, respectively. These energy bands were chosen to maximize the signal-to-noise ratio of the detection of the hot gas and the population of LMXBs.

For the soft-energy band, an optically-thin plasma emission is assumed, with a temperature of 0.2 keV and an iron abundance of 0.2 Z$_\odot$ to describe gaseous emission of UDGs. This assumption is motivated by the observed temperature and iron abundance of spiral and low-mass elliptical galaxies \citep[e.g.][]{goulding2016massive}. Although some individual galaxies could have different values for the gas temperature and iron abundance, it is unexpected that this variation affects our results, as we are studying a large number of galaxies. For broad-energy band, we assumed that the LMXB emission follows a power-law spectrum with an index of 1.7 \citep[e.g.][]{Irwin2003,piconcelli2005xmm}, as in \citet{kovacs2020ultra}.

\begin{figure*}
	\includegraphics[width=1.0\textwidth]{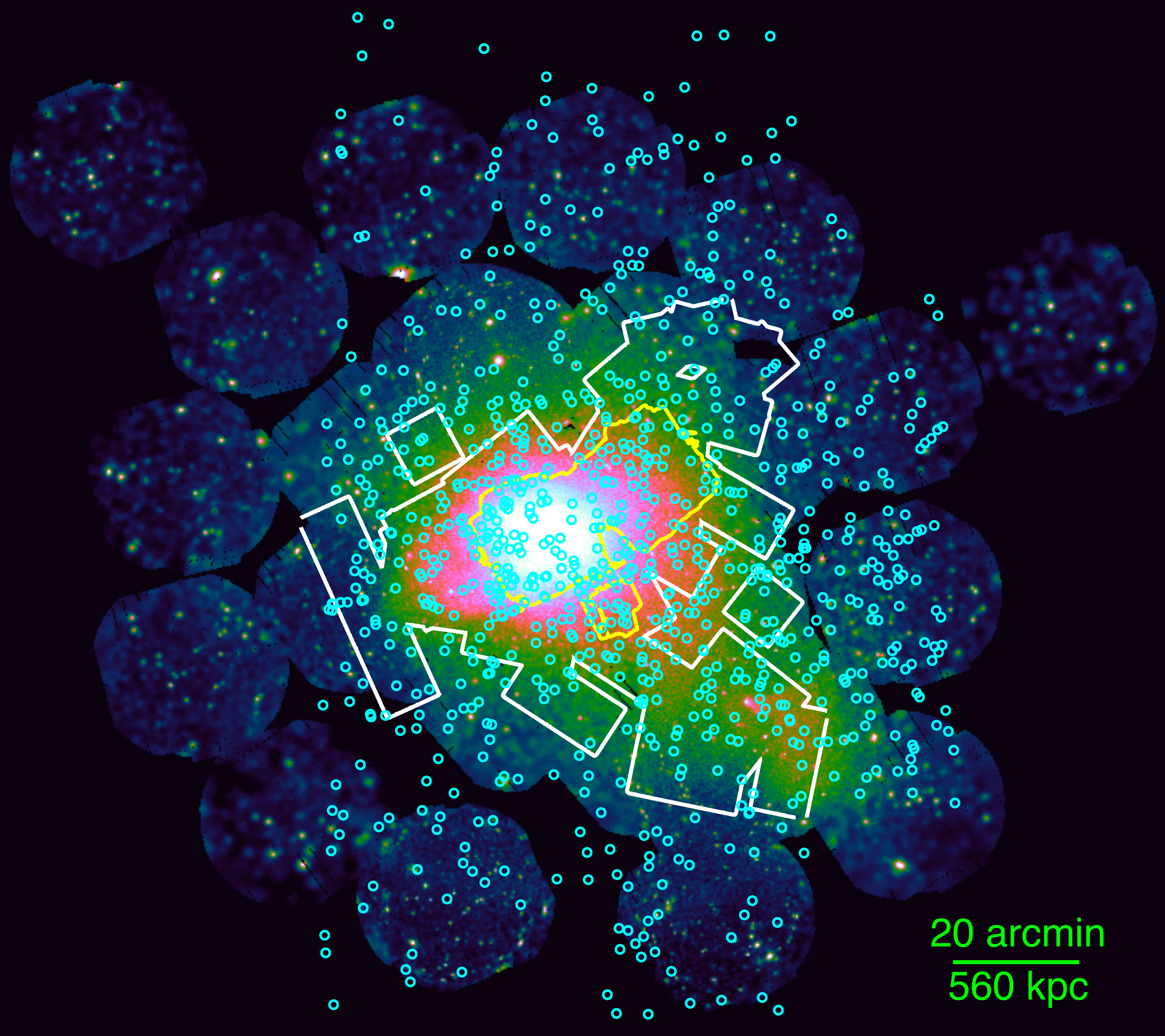}
	\caption{\textit{XMM$-$Newton} mosaicked image of the Coma cluster in the energy band 0.5$-$1.2 keV. The small cyan circles are centred on the UDG coordinates as given by Subaru survey \citep{yagi2016catalog}, and each set to have a radius of 40 arcsec to aid visibility. The white contours show the full extent of the \textit{Chandra} mosaic explored in \citet{kovacs2020ultra}. The bulk of the on-axis \textit{Chandra} exposure time is concentrated in 2 central pointings, shown by the yellow contours. By excluding the central 4 arcmin region, the \textit{XMM$-$Newton} observations are available for 779 out of 854 UDG candidates identified in the Coma cluster.}
	\label{fig: UDGs_Coma}
\end{figure*}

In Fig. \ref{fig: UDGs_Coma}, we show the \textit{XMM$-$Newton} mosaicked image of the Coma cluster in the energy band 0.5$-$1.2 keV. The small cyan circles ($r = 40$ arcsec) mark the locations of the UDG candidates in the Coma cluster detected in the Subaru survey \citep{yagi2016catalog}. The white contours show the full extent of the \textit{Chandra} mosaic explored in \citet{kovacs2020ultra}. The bulk of the on-axis \textit{Chandra} exposure time (around 1.6 Ms of the 2.0 Ms total) is concentrated in 2 central pointings, shown by the yellow contours. Our complete \textit{XMM--Newton} mosaic, with full coverage of Coma out to the virial radius in all directions, allows us to provide a near complete X-ray survey of the UDGs identified in the Subaru survey. By using a large mosaic of \textit{XMM$-$Newton} observations of the Coma cluster, we find that 786 UDG candidates have \textit{XMM$-$Newton} observations.

As in \citet{kovacs2020ultra}, we excluded the central 4 arcmin region of the Coma cluster from our analysis since the X-ray emission in this region is mainly dominated by the hot intracluster medium gas that distributed between galaxies \citep[e.g.][]{mirakhor2020complete}, which is significantly higher than the emission expected from UDGs. Beyond this radius, as the surface brightness of the intracluster medium drops significantly, it is expected the X-ray emission from various components of UDGs to play a more significant role. We find that there are only 7 UDG candidates in the central 4 arcmin region. By excluding these UDGs, the total number of UDGs analysed in this work reduces to 779 UDGs. This number makes about 91 per cent of the full sample of UDGs detected by Subaru survey \citep{yagi2016catalog}, and is almost twice the number of UDGs (404 UDGs) analysed in \citet{kovacs2020ultra} using \textit{Chandra} observations.

\section{X-ray analysis and Results}
\label{sec: results}

We probe the formation scenarios of UDGs in the Coma cluster using two approaches. In addition, we investigate the AGN occupation fraction of UDGs in the Coma cluster. As discussed in Section \ref{sec: intro}, we carry out a similar analysis to that described in \citet{kovacs2020ultra}, but, in the current work, the X-ray analysis is performed on 779 UDG candidates. Below, we present the approaches that used to probe the formation scenarios and AGN occupation fraction in the Coma cluster, and the results derived from the \textit{XMM$-$Newton} analysis.

\subsection{Dark matter halo mass}
\label{sec: halo_mass}
We first probed the formation scenarios of UDGs by measuring the X-ray luminosity of their hot halos in the soft energy band. As the X-ray luminosity of the gaseous X-ray halos is a robust tracer of the total gravitating mass of galaxies \citep[e.g.][]{kim2013x,babyk2018x,lakhchaura2018thermodynamic}, measuring the X-ray luminosity of the UDG hot halos allows to set a proper constraint on the dark matter halo mass of UDGs, and therefore their formation mechanism. UDGs could be the descendants of massive galaxies if they reside in massive dark matter halos similar to halos of Milky Way-type galaxies. Alternatively, UDGs could be genuine dwarf galaxies with large spatial extent if they reside in low-mass dark matter halos.    

To test whether UDGs reside in massive or low-mass dark matter halos, \citet{kovacs2020ultra} estimated the expected luminosity of the gaseous X-ray halo around a UDG, assuming that it resides in a Milky Way-type dark matter halo with a virial mass of $M_{\rm{vir}} = 8.0 \times 10^{11}$ M$_{\odot}$. Within a radius of $5r_{\rm{eff}}$, this virial mass corresponds to a total mass of $M_{\rm{tot}} = 1.8 \times 10^{11}$ M$_{\odot}$. Using the $L_{0.3-8{\rm{keV}}}\!-\!M_{\rm{tot}}$ scaling relation \citep{kim2013x}, the expected X-ray luminosity in the 0.3$-$8.0 keV band estimated by these authors is $\approx 1.8 \times 10^{39}$ erg s$^{-1}$. Assuming a temperature of 0.2 keV and an iron abundance of 0.2 Z$_\odot$, the corresponding value for the expected luminosity in the 0.5$-$1.2 keV band is $\approx 1.1 \times 10^{39}$ erg s$^{-1}$. This value is relatively small and falls well below the X-ray luminosity of most individual galaxies detected by \textit{XMM-Newton}. To overcome this limitation, \citet{kovacs2020ultra} stacked the X-ray emission originating from individual galaxies, allowing to probe the emission of the average UDG population with better sensitivity.

\begin{figure}
	\includegraphics[width=\columnwidth]{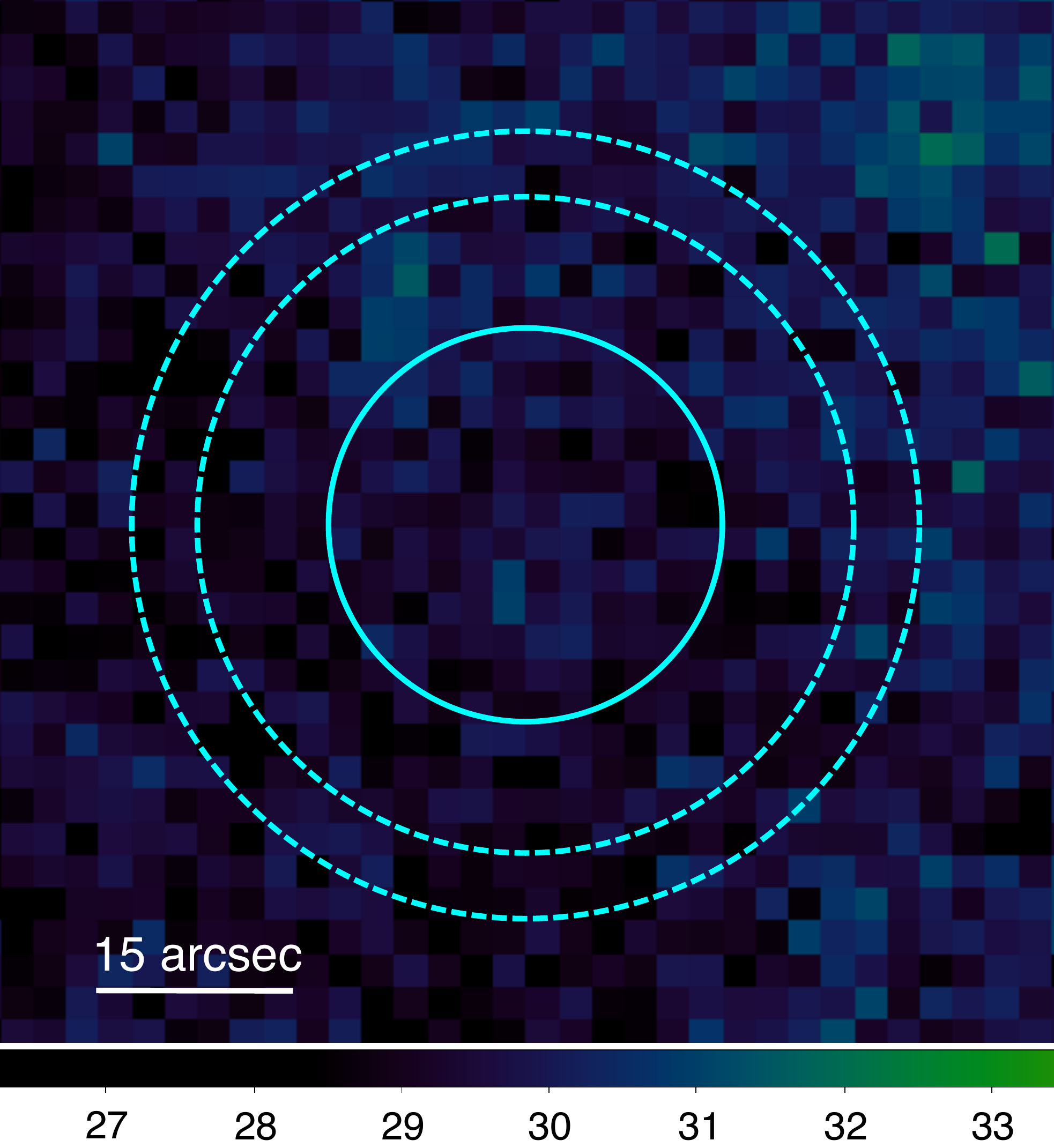}
	\caption{Stacked exposure-corrected image of the analysed Coma cluster UDGs in the soft band. The stacked image centred on the UDG coordinates as given by Subaru survey \citep{yagi2016catalog}. The cyan circle (solid line) with 15 arcsec radius marks the region where the source counts were extracted. The circular annuli (dashed lines) with radii of 25$-$30 arcsec marks the region in which the background counts were extracted. We did not obtain significant X-ray emission from the hot halo of the analysed UDGs, suggesting that the bulk of the UDG population in the Coma cluster are genuine dwarf galaxies.}
	\label{fig: stacked_image}
\end{figure}

Following \citet{kovacs2020ultra}, we performed the stacking analysis in the soft band (0.5$-$1.2 keV) to measure the gaseous X-ray luminosity of the UDG population. For each UDG in our sample, we extracted a region of 150 arcsec $\times$ 150 arcsec from the X-ray image and exposure map, centred on the UDG position as given by Subaru survey. We then combined the extracted images and exposure maps for the full list of UDGs in our sample into a single image. Fig. \ref{fig: stacked_image} shows the stacked exposure-corrected image of the analysed UDGs in the soft band. \citet{kovacs2020ultra} measured the X-ray luminosity associated with the UDG population within a circular region of 5 arcsec radius. However, as \textit{XMM--Newton} has a broader PSF than \textit{Chandra}, the X-ray counts associated with the UDGs may be scattered to a much larger region. Within a circular region of 5 arcsec radius, only 35 per cent of the PSF is encircled within this region. We therefore considered a larger extraction region for measuring the X-ray luminosity. Motivated by the PSF size of \textit{XMM--Newton}, we extracted the source counts using a circular region of 15 arcsec radius, as we find that 90 per cent of the PSF is enclosed within this region. The background counts were extracted from an annulus region with 25$-$30 arcsec radii.

After accounting for the stacked source and background counts, we did not obtain statistically significant X-ray emission from the hot halo of the analysed UDGs, agreeing with that reported in \citet{kovacs2020ultra}. As a result of the absence of significant X-ray emission, a 2$\sigma$ upper limit on the X-ray luminosity is computed. Assuming an optically-thin plasma emission with a temperature of 0.2 keV and an iron abundance of 0.2 Z$_\odot$, we obtain a 2$\sigma$ upper limit of $< 1.0 \times 10^{-16}$ erg s$^{-1}$ cm$^{-2}$ on the average X-ray flux of the analysed Coma cluster UDGs in the 0.5$-$1.2 keV energy range. At the redshift of Coma, this value places a 2$\sigma$ upper limit of $< 1.2 \times 10^{38}$ erg s$^{-1}$ on the X-ray luminosity of the hot gaseous halo. This upper-limit luminosity, although slightly higher than that estimated for the analysed Coma cluster UDGs in \citet{kovacs2020ultra}, is about 9 times lower than the luminosity of $1.1 \times 10^{39}$ erg s$^{-1}$ estimated for a galaxy with a massive dark matter halo. These results imply that significant fraction of UDGs in the Coma cluster are genuine dwarf galaxies, in good agreement with the results obtained for the Coma cluster UDGs studied in \citet{kovacs2020ultra}.  

\begin{figure}
	\includegraphics[width=\columnwidth]{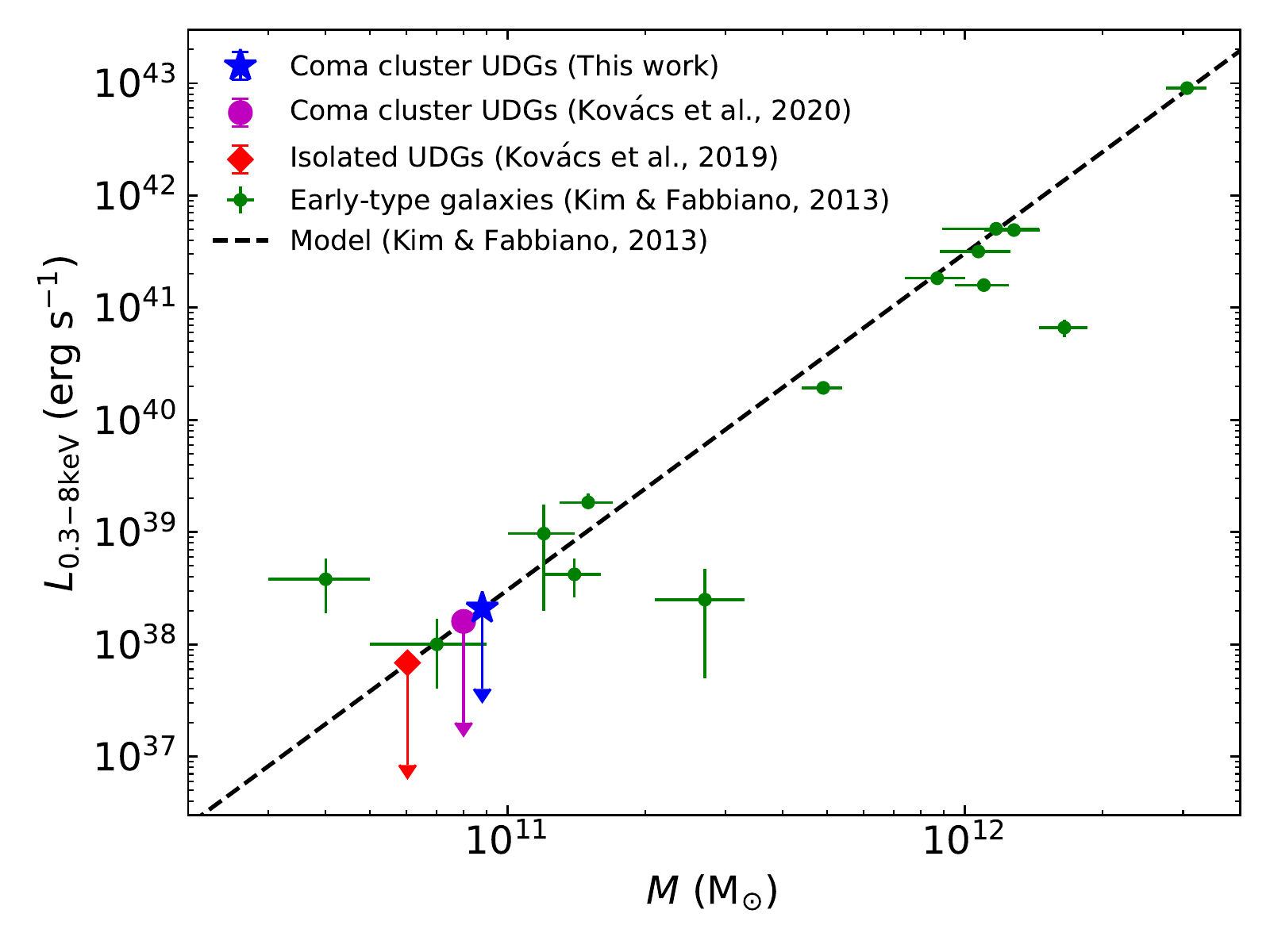}
	\caption{Correlation between the X-ray luminosity of hot gas in the 0.3$-$8.0 keV band and the total gravitating mass for a sample of early-type galaxies \citep{kim2013x}, isolated UDGs \citep{kovacs2019constraining}, Coma cluster UDGs \citep{kovacs2020ultra}, and Coma cluster UDGs analysed in this work. Using the best-fitting relation \citep{kim2013x}, the 2$\sigma$ upper limit on the X-ray luminosity implies a total mass of $< 9.0 \times 10^{10}$ M$_{\odot}$ within $5 r_{\rm{eff}}$. Although this mass is slightly higher than a mass of $< 8.0 \times 10^{10}$ M$_{\odot}$ estimated for the analysed Coma cluster UDGs in \citet{kovacs2020ultra}, it is only half the total mass obtained for a galaxy with a massive dark matter halo. This suggests that the bulk of UDGs in the Coma cluster reside in low-mass dark matter halos.}
	\label{fig: L_M_correlation}
\end{figure}

The 2$\sigma$ upper-limit luminosity of the hot gaseous halo in the soft band corresponds to an X-ray luminosity of $ < 2.1 \times 10^{38}$ erg s$^{-1}$ in the 0.3$-$8.0 keV band. Using the best-fitting relation that found between the gas luminosity and total mass \citep{kim2013x}, this value implies a total mass of $M_{\rm{tot}} < 9.0 \times 10^{10}$ M$_{\odot}$ within $5 r_{\rm{eff}}$. This mass is only half the total mass obtained for a galaxy with a Milky Way-type dark matter halo. In Fig. \ref{fig: L_M_correlation}, we show the correlation between the X-ray luminosity of hot gas in the 0.3$-$8.0 keV band and the total gravitating mass for a sample of early-type galaxies \citep{kim2013x}, isolated UDGs \citep{kovacs2019constraining}, Coma cluster UDGs \citep{kovacs2020ultra}, and Coma cluster UDGs analysed in this work. In the same figure, we also show the best-fitting relation between the X-ray luminosity and the total gravitating mass obtained from the X-ray scaling relation in early-type galaxies \citep{kim2013x}.

\subsection{GC-LMXB population}
\label{sec: GC_LMXB}
We also probe the origin of UDGs in the Coma cluster through their GC-LMXB population, as is done in \citet{kovacs2020ultra}. Due to the high-stellar density in the core, GCs can effectively trigger the formation of various types of binaries, including LMXBs \citep[e.g.][]{kim2006low,sivakoff2007low}. It is well known from X-ray studies \citep[e.g.][]{sarazin2000resolving} that the bulk of the X-ray emission in early-type galaxies originates from LMXBs. Also, it is well established that the number of LMXBs per unit stellar mass is a few hundred times higher in GCs than in the galactic field, as stellar densities in the field are typically very low to form LMXBs \citep[e.g.][]{angelini2001x,minniti2004optical,jordan2007low}. Therefore, we do not expect to see significant amount of the X-ray emission from field LMXBs. On the other hand, if UDGs reside in massive dark matter halos and host large numbers of GCs as found in two Coma cluster UDGs: Dragonfly 44 and DF X1 \citep{van2017extensive}, significant amount of the X-ray emission is then expected from GC-LMXBs. By comparing the X-ray luminosity of GC-LMXBs with the luminosity expected from a galaxy with a massive dark matter halo, one can constrain whether the bulk of the UDG population hosts a large number of GCs. Consequently, this allows us to determine whether UDGs reside in massive or low-mass dark matter halos.

As discussed in Section \ref{sec: halo_mass}, if we assume that UDGs reside in massive dark matter halos with a virial mass of $M_{\rm{vir}} = 8.0 \times 10^{11}$ M$_{\odot}$, and based on the relation between the number of GCs and the virial mass of the host galaxy \citep{burkert2020high}, each UDG should then host, on average, 160 GCs. Using the average GC-LMXB luminosity function, this number of GCs, in turn, implies a combined GC-LMXB luminosity of about $9.0 \times 10^{39}$ erg s$^{-1}$ in the 0.5$-$7.0 keV band.

Using a sample of galaxies, \citet{zhang2011luminosity} found that the X-ray luminosity for a large number of GC-LMXBs is less than $10^{38}$ erg s$^{-1}$, suggesting that most of them remain unresolved at the redshift of Coma based on the existing \textit{XMM$-$Newton} observations. Therefore, we carried out the stacking analysis, as is done in Section \ref{sec: halo_mass}, but this time in the 0.5$-$7.0 keV energy range, by co-adding the X-ray images and exposure maps associated with the individual UDGs. The source counts were extracted from a circular radius of 15 arcsec, and the background counts were extracted from a circular annulus with radii of 25$-$30 arcsec. As in Section \ref{sec: halo_mass}, we did not obtain statistically significant X-ray emission from the source region, in agreement with that reported in \citet{kovacs2020ultra}. In the absence of significant emission, we computed the $2\sigma$ upper-limit flux by assuming that the X-ray emission from GC-LMXBs follows a power-law spectrum with a spectral index of 1.7. The 2$\sigma$ upper-limit flux is $< 1.4 \times 10^{-16}$ erg s$^{-1}$ cm$^{-2}$, which places a $2\sigma$ upper limit of $< 1.8 \times 10^{38}$ erg s$^{-1}$ on the X-ray luminosity, at the redshift of the Coma cluster. 

Despite that the upper limit on the X-ray luminosity is higher than the $2\sigma$ upper-limit luminosity of $< 1.1 \times 10^{38}$ erg s$^{-1}$ reported in \citet{kovacs2020ultra}, it is  about 50 times lower than that predicted for a UDG hosting a large number of GCs. This result indicates that the bulk of the UDG population in the Coma cluster does not host a large number of GCs, agreeing with that found in \citet{kovacs2020ultra}. This implies that significant fraction of the analysed UDGs are not the descendants of massive galaxies, but are genuine dwarf galaxies.


\subsection{X-ray AGN identification}
\label{sec: AGN_search}
Since the mass of AGN correlates with the dark matter halo mass of the host galaxy \citep[e.g.][]{bogdan2015connecting}, it is interesting to test whether the AGN occupation fraction of UDGs is comparable with that obtained for massive or dwarf galaxies. \citet{kovacs2020ultra} reported only two UDGs out of 404 that have an off-centre X-ray source within a radius of 5 arcsec from the centre of the UDGs. Thus, they placed an upper limit of $< 0.5$ per cent on the occupation fraction of AGN in the Coma cluster UDGs. This value falls short of that reported for high-mass galaxies, but is more comparable with that found for dwarf galaxies \citep[e.g.][]{miller2015x,kaviraj2019agn}. 

\begin{figure*}
	\includegraphics[width=1.0\textwidth]{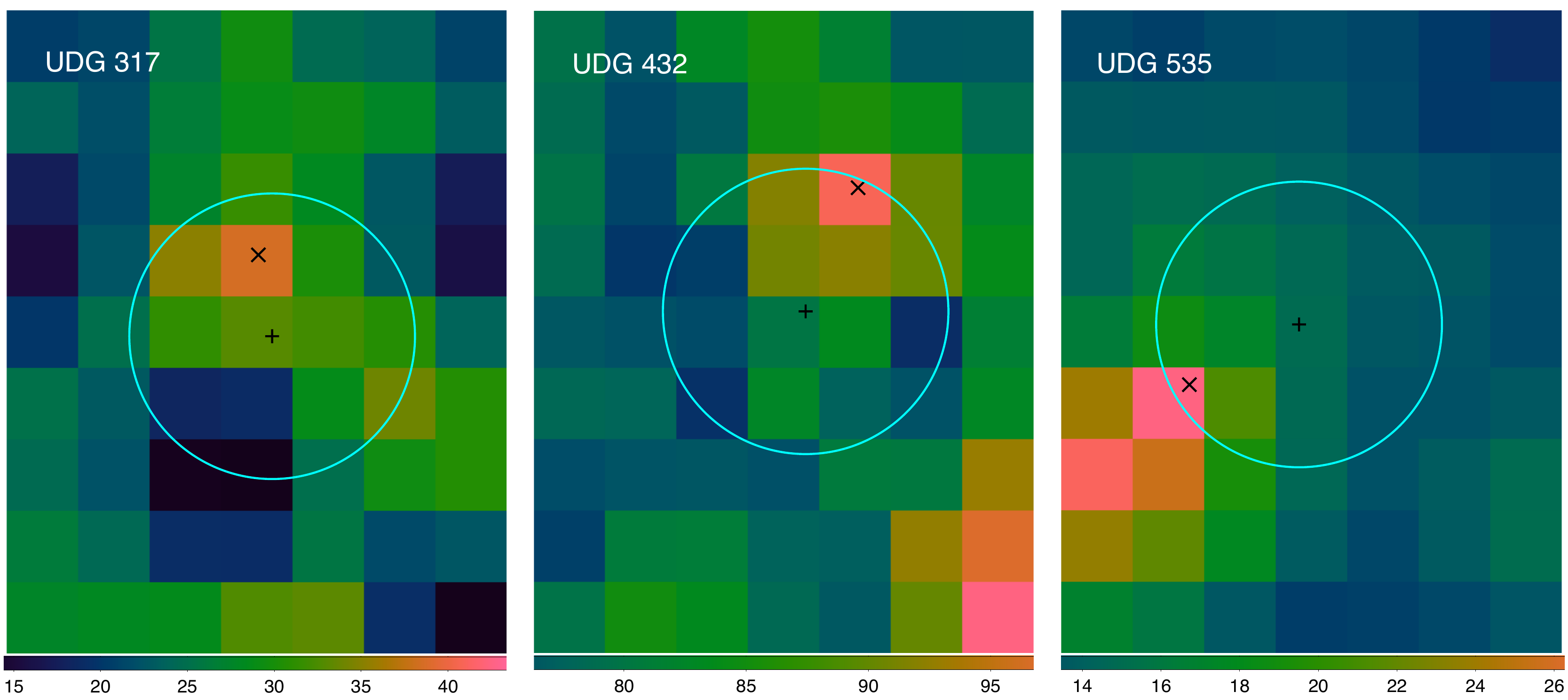}
	\caption{Exposure-corrected \textit{XMM$-$Newton} images of UDG 317, UDG 432, and UDG 535 in the energy range 0.5$-$1.2 keV. The cyan circles have an angular radius of 5 arcsec and are centred on the UDG coordinates, shown as black plus marks. The coordinates of the detected point sources that lie within 5 arcsec from the UDG centres are shown as black cross marks.}
	\label{fig: UDG_matches}
\end{figure*}

 In this work, the AGN occupation fraction is computed using an approach similar to that presented in \citet{kovacs2020ultra}. We searched for AGN in the Coma cluster UDGs by cross-correlating the position of the analysed UDGs with the position of the point sources detected in the soft and broad bands. As in \citet{kovacs2020ultra}, we did not find any matches within a search radius of 2.5 arcsec from the centre of the UDGs. Increasing the search radius to 5 arcsec (corresponding to $2r_{\rm{eff}}$ at the redshift of Coma), we detected three X-ray sources with detection significance $\sigma \geq 5$, in which their position matches with the position of the Coma's UDGs. These galaxies are the UDG 317 ($r_{\rm{eff}}=2.7$ arcsec), UDG 432 ($r_{\rm{eff}}=1.9$ arcsec), and UDG 535 ($r_{\rm{eff}}=3.2$ arcsec), and the radii of the X-ray sources associated with them are 4.3, 5.1, and 7.3 arcsec, respectively. The offsets between these UDG coordinates and the detected point sources are, respectively, 2.9, 4.7, and 4.4 arcsec for the UDG 317, UDG 432, and UDG 535. In Fig. \ref{fig: UDG_matches}, we show the X-ray images of these UDGs and the detected point sources.   

The identified point source within 5 arcsec from the UDG 317 centre (Fig. \ref{fig: UDG_matches}, left panel) is reported by \citet{kovacs2020ultra}. These authors also identified an X-ray source around UDG 412. However, we did not detect any X-ray sources around UDG 412 using the same searching radius. On the other hand, the identified X-ray sources around UDG 432 and UDG 535 (Fig. \ref{fig: UDG_matches}, middle and right panels) are not reported by \citet{kovacs2020ultra}. This is expected since the number of the studied UDGs here is greater, by a factor of about 2, than that studied in \citet{kovacs2020ultra}.

If we assume that these X-ray sources reside in the Coma cluster, their X-ray luminosity in the 0.5$-$7.0 energy band is $(8.1 \pm 0.8) \times 10^{38}$ erg s$^{-1}$, $(2.1 \pm 0.2) \times 10^{39}$ erg s$^{-1}$, and $(5.5 \pm 0.6) \times 10^{38}$ erg s$^{-1}$ for the UDG 317, UDG 432, and UDG 535, respectively. The luminosity of the identified point sources in the UDG 317 and UDG 535 is below $10^{39}$ erg s$^{-1}$, and their $B-R$ color indices are 0.91 and 0.93, respectively. Therefore, it is unlikely that  these sources are high-mass X-ray binary (HMXB) or ultra-luminous X-ray (ULX) sources, but are more likely LMXBs \citep[e.g.][]{gilfanov2004low,kim2004x}. It is also possible that these sources are off-centre low-luminosity AGN. However, with a luminosity below $10^{39}$ erg s$^{-1}$, it is difficult to distinguish between the X-ray emission originating from an AGN and X-ray binaries \citep[e.g.][]{lehmer2010chandra,baldassare2018identifying}. For the UDG 432, the luminosity of the identified point source exceeds $10^{39}$ erg s$^{-1}$, indicating that it could either be a HMXB or ULX source. These X-ray sources, however, are typically associated with the ongoing star-formation activity \citep[e.g.][]{mineo2012x,sazonov2017bright}. It is found that the UDG 432 has a $B-R$ color index of 1.1 with no signature of H$\alpha$ emission \citep{koda2015approximately}, implying that this galaxy is not forming stars at the current epoch. This, in turn, suggests that the associated X-ray source is unlikely to be an HMXB or ULX, but is more likely an off-centre low-luminosity AGN, assuming that the UDG 432 resides in the Coma cluster and the X-ray source lies within the galaxy. However, there is still a possibility that the X-ray source is not a member of the Coma cluster, and is therefore a foreground or background object. 

As the number of counts collected from the X-ray sources is very low, it is not possible to carry out precise spectroscopic measurements. However, as is done in \citet{kovacs2020ultra}, we computed a hardness ratio for these sources to determine their nature, using counts in the hard and soft energy bands. The hardness ratio is defined as $HR=(H-S)/(H+S)$, where $H$ and $S$ are the net counts in the hard and soft bands, respectively. The computed hardness ratio for the X-ray sources is 0.19, 0.22, and 0.20 for the UDG 317, UDG 432, and UDG 535, respectively. Assuming the X-ray spectrum of AGN follows a power law with a spectral slope lying in the range of the typically observed values ($\Gamma =1.5\!-\!2.0$), the corresponding value for the AGN hardness ratio is then in the 0.04$-$0.23 range. This implies that the X-ray sources in the UDG 317, UDG 432, and UDG 535 might originate from AGN. A similar conclusion is obtained for the X-ray sources identified in \citet{kovacs2020ultra}.          

With the current X-ray data, however, it is difficult to conclusively determine the true nature of the X-ray sources associated with these galaxies. Since we cannot confirm the nature of these sources, we set an upper limit on the AGN occupation fraction of UDGs. Based on the identification of three point sources within 5 arcsec from the UDG centre, which could be off-centre AGN, we obtain an upper limit of < 0.4 per cent on the AGN occupation fraction of the Coma cluster UDGs. This is in agreement with that reported in \citet{kovacs2020ultra}.

\section{Discussion}
\label{sec: discussion}

\subsection{UDG origin}
\label{sec: UDG_origin}
We have investigated the formation scenarios of UDGs in the Coma cluster using two different approaches. The analysis procedure carried out in this work is similar to that used in \citet{kovacs2020ultra}. However, we have used a larger population of UDGs, which is almost twice the population of UDGs used in \citet{kovacs2020ultra}.  

By utilizing \textit{XMM$-$Newton} observations, we have probed the formation scenarios of UDGs in the Coma cluster through their hot gaseous halos and LMXBs residing in GCs. We did not obtain statistically significant X-ray emission from the hot gaseous halos or from GC-LMXBs in the stacked sample. In the absence of significant emission, we set a $2\sigma$ upper limit on measurements. Assuming an optically-thin plasma emission with a temperature of 0.2 keV and a metallicity of 0.2 Z$_\odot$, we have obtained a $2\sigma$ upper limit of $< 1.2 \times 10^{38}$ erg s$^{-1}$ on the X-ray luminosity of the hot gaseous halo, which is around 9 times lower than that derived for a galaxy residing in a massive dark matter halo. For GC-LMXBs, we have obtained a $2\sigma$ upper limit of $< 1.8 \times 10^{38}$ erg s$^{-1}$ on the X-ray luminosity, which is around 50 lower than that derived for a galaxy hosting a large number of GCs. These results indicate that the average UDG population does not have a substantial amount of hot X-ray gas or a large number of GCs. Based on these results, we conclude that most UDGs in the Coma cluster are genuine dwarf galaxies. This conclusion agrees with that obtained for the Coma cluster UDGs studied in \citet{kovacs2020ultra} using \textit{Chandra} observations. It is also in agreement with the conclusion derived for a sample of isolated UDGs \citep{kovacs2019constraining}.

However, we cannot rule out the possibility that some individual UDGs, or small sub-samples of them, may have different formation pathways. It is possible that some UDGs are the result of processing by the large-scale dense environment, and formed either as extended, low-surface brightness galaxies or as small, low-mass galaxies \citep[e.g.][]{gnedin2003dynamical,collins2013kinematic}. It is also possible that some UDGs are the descendants of massive galaxies, but lost their gas at high redshift due to extreme feedback from ram-pressure stripping \citep[e.g.][]{fujita2004pre}, supernovae \citep[e.g.][]{agertz2016impact}, and AGN activity \citep[e.g.][]{moran2014black}. This, in turn, quenched the star formation in these systems, forming extended and low-surface brightness galaxies with massive dark matter halos. 

\citet{beasley2016overmassive} found that the UDG VCC 1287 in the Virgo cluster hosts a large GC population. Similar results were also reported for the UDGs Dragonfly 17 \citep{peng2016rich}, Dragonfly 44, and DF X1 \citep{van2017extensive} in the Coma cluster. The number of GCs detected in these systems is significantly larger than that typically detected in dwarf galaxies \citep[e.g.][]{georgiev2010globular}. This large GC population around these galaxies supports the formation scenario, in which UDGs reside in massive dark matter halos. These results, however, do not contradict with our findings since we stacked a large number of UDGs, and we therefore probed their average properties. Besides, more recent studies \citep{lee2020ultraviolet,bogdan2020archetypal} have found that Dragonfly 44 and DF X1 do not host significant X-ray emission. Therefore, as for the bulk of the UDGs, the same conclusion can also be applied to these galaxies.

\subsection{AGN occupation fraction}
\label{sec: AGN_fraction}
In addition to probing the formation scenarios, we have constrained the AGN occupation fraction of UDGs in the Coma cluster. As discussed in Section \ref{sec: AGN_search}, we have identified three X-ray sources within a radius of 5 arcsec from the centre of the UDG 317, UDG 432, and UDG 535 (Fig. \ref{fig: UDG_matches}). Based on the current data, we cannot conclusively determine the redshift and nature of the X-ray sources associated with these UDGs. They could be off-centre AGN residing in the galaxies, foreground X-ray binaries, or background AGN.   

However, if we assume that these UDGs reside in the Coma cluster and the identified X-ray sources are within the field of these galaxies, these sources are then likely off-center low-luminosity AGN. The offset between these UDG coordinates and the detected X-ray sources ranges between 2.9$-$4.7 arcsec, which corresponds to 1.4$-$2.3 kpc at the redshift of the Coma cluster. The presence of such offsets is likely in dwarf galaxies, and is in agreement with those reported in \citet{kovacs2020ultra}. These offsets are also in agreement with those offsets found in X-ray and radio surveys of AGN in dwarf galaxies \citep[e.g.][]{mezcua2018intermediate,reines2020new}. Furthermore, high-resolution cosmological simulations \citep[e.g.][]{bellovary2019multimessenger} predicted that about half of massive black holes in dwarf galaxies are not centrally located, but rather are locating at distances of a few kpc from the galaxy centre. 

Based on the identification of three X-ray sources within 5 arcsec from the galaxy centre, which could be off-centre low-luminosity AGN, we have placed an upper limit of < 0.4 per cent on the occupation fraction of AGN in the Coma cluster UDGs. This upper limit agrees well with an upper limit of < 0.5 per cent on the AGN occupation fraction reported in \citet{kovacs2020ultra}. Our result is also consistent with an AGN occupation fraction of about 0.6 per cent reported for a sample of dwarf galaxies with stellar masses of $10^{8.5} < M_{\rm{star}} < 10^{9.5}$ M$_{\rm{\odot}}$ utilizing data from the Sloan Digital Sky Survey \citep[SDSS;][]{reines2013dwarf}. However, our estimated value for the AGN occupation fraction falls significantly short of the values obtained for early-type galaxies in the AMUSE (AGN Multi-wavelength Survey of Early-type galaxies) survey \citep{miller2015constraining}. For galaxies with $M_{\rm{star}} < 10^{10}$ M$_{\rm{\odot}}$, these authors placed an lower limit of 20 per cent on the AGN occupation fraction, which exceeds the upper limit obtained for the UDGs in the Coma cluster by a factor of about 50. This difference in the AGN occupation fraction may be partially due to the higher sensitivity of the AMUSE survey, which allows to detect fainter X-ray sources.


It is possible that the X-ray sources associated with the UDG 317, UDG 432, and UDG 535 are not AGN in these galaxies, but are foreground or background objects. Therefore, their existence within a radius of 5 arcsec from the centre of the UDGs is due to random matches. To quantify the likelihood of random matches, we performed Monte Carlo simulations, as is done in \citet{kovacs2020ultra}. After excluding the central 4 arcmin region, we generated 779 random coordinates within the footprint of the \textit{XMM--Newton} observations for the Coma cluster. We then searched for coincident matches within 5 arcsec between the position of the randomly generated coordinates and the position of the detected X-ray sources in the soft and broad bands. We repeated this process $10^5$ times, and the number of random matches was recorded at each time.

\begin{figure}
	\includegraphics[width=\columnwidth]{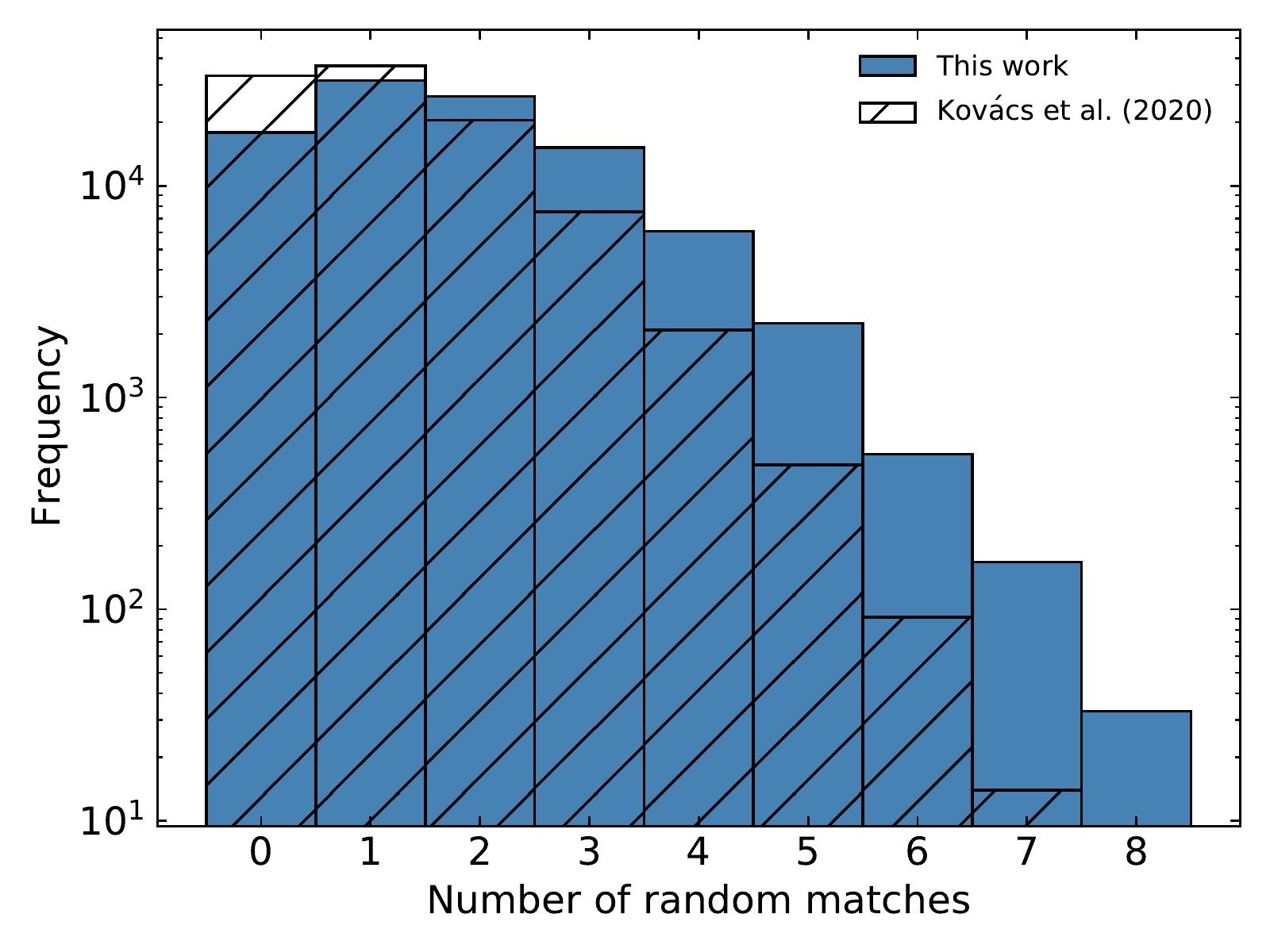}
	\caption{Number of random X-ray source-UDG matches within the \textit{XMM--Newton} footprint of the Coma cluster for 779 randomly generated UDG coordinates using Monte Carlo simulations with 10$^5$ trials. We also show the result of Monte Carlo simulations from \citet{kovacs2020ultra} (hatched histogram). The number of random matches ranges between 0$-$8, with about 60 per cent of them in the range of 1 to 2. The average number of random matches is 1.70, suggesting that two of the three identified X-ray sources, at least, are likely due to the result of random matches.}
	\label{fig: histogram}
\end{figure}

In Fig. \ref{fig: histogram}, we present a histogram showing the result of Monte Carlo simulations. The number of random X-ray source-UDG matches ranges between 0$-$8, with about 60 per cent of them in the range of 1 to 2. The average number of random matches is 1.70, which corresponds to an average AGN occupation fraction of 0.22 per cent. These results suggest that two of the three identified X-ray sources, at least, are likely due to the result of random matches.      

To set further constraint on the nature of the identified X-ray sources, we inspected the regions around the UDG 317, UDG 432, and UDG 535 in the optical band using Digitized Sky Survey (DSS) and Subaru images. We identified an optical source within a radius of 5 arcsec from the UDG 317 centre, matching the position of the X-ray source. However, no optical counterparts were detected for the X-ray sources associated with the UDG 432 and UDG 535. Also, we checked images of these UDGs in FIR and other wave bands, but we did not find any significant sources in the field of these galaxies. Using the Subaru images, \citet{kovacs2020ultra} also identified an optical counterpart for the X-ray source associated with the UDG 317. The detection of an optical counterpart implies that the X-ray source associated with the UDG 317 is an off-centre AGN. However, this detection does not constrain whether the detected AGN is residing in the Coma cluster or not. Therefore, we cannot rule out the possibility that it could be a background or foreground object.

\section{Conclusions}
In this work, we have probed the formation scenarios and AGN occupation fraction of 779 UDG candidates in the nearby Coma cluster, following an X-ray analysis similar to that presented in \citet{kovacs2020ultra}. The formation scenarios of UDGs were probed by constraining their dark matter halo mass using two approaches. Since the dark matter halo mass correlates with the X-ray luminosity, we have measured the X-ray luminosity of hot gas and GC-LMXB by stacking the X-ray photons of a large sample of UDGs. We did not detect statistically significant emission from hot gas or from LMXBs residing in GCs. In the absence of significant detection, we have placed $2\sigma$ upper limits of $< 1.2 \times 10^{38}$ erg s$^{-1}$ and $< 1.8 \times 10^{38}$ erg s$^{-1}$ on the X-ray luminosity expected from hot gas and GC-LMXBs, respectively. The upper limits fall significantly short of the X-ray luminosities expected from UDGs residing in massive dark matter halos and hosting significant population of GCs. These results suggest that significant fraction of UDG candidates in the Coma cluster reside in low-mass dark matter halos, implying that they are puffed-up dwarf galaxies. This conclusion agrees with that obtained for 404 UDGs in the Coma cluster using \textit{Chandra} observations \citep{kovacs2020ultra}. It is also in agreement with the conclusion derived for a sample of isolated UDGs \citep{kovacs2019constraining}.  

Moreover, we have constrained the AGN occupation fraction of UDGs by cross-correlating the position of UDGs with the identified X-ray sources in the Coma cluster. We have identified three X-ray sources that could be AGN at distances of 2.9, 4.7, and 4.4 arcsec from the centre of the UDG 317, UDG 432, and UDG 535, respectively. Since spectroscopic measurements are not available for them, we cannot conclusively determine the true nature of these sources. Also, Monte Carlo simulations indicate that two of these sources are likely due to spatial coincidence. Based on these results, we have placed an upper limit of < 0.4 per cent on the AGN occupation fraction of UDGs in the Coma cluster, which is comparable with that found for dwarf galaxies.

Deep observations with the current X-ray observatories are required to carry out precise spectroscopic measurements and establish whether the UDGs and associated X-ray sources are residing in the Coma cluster. However, a complete understanding of the origin and evolutionary picture of UDGs can only be achieved with upcoming high-resolution spectral X-ray observations such as \textit{Athena} \citep{ATHENA}, and proposed concepts such as \textit{Lynx} \citep{Lynx} and \textit{AXIS} \citep{AXIS}.

\section*{Acknowledgements}
We thank the referee for their helpful report. MSM and SAW
acknowledge support from the NASA \textit{XMM--Newton} grant 19-XMMNC18-0030. Based on observations obtained with \textit{XMM--Newton}, an ESA science mission with instruments and contributions directly funded by ESA Member States and NASA.

\section*{Data Availability}
The \textit{XMM--Newton} Science Archive (XSA) stores the archival data used in this paper, from which the data are publicly available for download.  The \textit{XMM} data were processed using the \textit{XMM--Newton} Science Analysis System (SAS). The
software packages \textsc{heasoft} and \textsc{xspec} were used, and these can be downloaded from the High Energy Astrophysics Science Archive Research Centre (HEASARC) software web-page. Analysis and figures were produced using \textsc{python} version 3.7.



\bibliographystyle{mnras}
\bibliography{UDG} 




\appendix

\section{\textit{XMM--Newton} observations of the Coma cluster}

\begin{table*}
    \centering
    \caption{\textit{XMM$-$Newton} observations of Coma}
    \begin{tabular}{lccccc}
    \hline
    
   Observation & Obs. ID & Obs. Date & RA & Dec. & Exposure \\
     &   &  & (J2000) & (J2000)   & (ks)  \\
    \hline
    Outskirts1    & 0841680101 & 07 Jul 2019 & 12 55 25.59 & +27 47 01.6 & 23.0 \\ 
    Outskirts2    & 0841680201 & 07 Jul 2019 & 12 55 38.07 & +28 17 43.1 & 23.0 \\
    Background1   & 0841681101 & 12 Jul 2019 & 12 53 31.92 & +28 29 54.4 & 29.5 \\
    Outskirts3    & 0841680301 & 11 Jul 2019 & 12 57 03.20 & +28 41 26.5 & 26.0 \\
    Outskirts4    & 0841680401 & 13 Jul 2019 & 12 59 11.76 & +28 52 56.4 & 23.0 \\
    Outskirts5    & 0841680501 & 13 Jul 2019 & 13 01 27.90 & +28 51 36.7 & 23.0 \\
    Outskirts6    & 0841680601 & 17 Jul 2019 & 13 03 15.79 & +28 32 54.7 & 23.0 \\
    Background2   & 0841681201 & 29 Dec 2019 & 13 05 10.70 & +28 53 08.8 & 23.0 \\
    Outskirts7    & 0841680701 & 16 Jul 2019 & 13 04 03.83 & +28 04 45.1 & 26.4 \\
    Outskirts8    & 0841680801 & 17 Jul 2019 & 13 03 56.84 & +27 35 23.5 & 34.4 \\
    Outskirts9    & 0841680901 & 06 Dec 2019 & 13 02 49.88 & +27 11 18.6 & 30.8 \\
    Outskirts10   & 0800580101 & 23 Dec 2017 & 13 00 57.97 & +26 58 35.3 & 80.4 \\
    Outskirts10   & 0800580201 & 04 Jan 2018 & 13 00 57.97 & +26 58 35.3 & 87.0 \\
    Outskirts11   & 0403150301 & 17 Jun 2006 & 12 57 40.75 & +26 56 13.6 & 56.0 \\
    Outskirts11   & 0403150401 & 21 Jun 2006 & 12 57 40.75 & +26 56 13.6 & 64.4 \\
    Outskirts12   & 0058940701 & 10 Jun 2003 & 12 55 24.99 & +27 13 46.0 & 22.6 \\
    Inside1         & 0124710501 & 29 May 2000 & 12 59 27.49 & +27 46 53.0 & 29.9 \\
    Inside2        & 0124711401 & 29 May 2000 & 12 59 46.71 & +27 56 60.0 & 34.6 \\
    Inside3         & 0124711601 & 11 Jun 2000 & 12 57 42.51 & +27 43 38.0 & 87.9 \\
    Inside4         & 0124710201 & 11 Jun 2000 & 12 57 42.51 & +27 43 38.0 & 41.5 \\
    Inside5         & 0124710901 & 11 Jun 2000 & 13 00 32.68 & +27 56 59.0 & 31.2 \\
    Inside6         & 0124710601 & 12 Jun 2000 & 12 58 50.01 & +27 58 52.0 & 31.8 \\
    Inside7         & 0124710101 & 21 Jun 2000 & 12 56 47.68 & +27 24 07.0 & 41.5 \\
    Inside8         & 0124710401 & 23 Jun 2000 & 13 00 04.60 & +27 31 24.0 & 52.8 \\
    Inside9         & 0124711101 & 24 Jun 2000 & 12 58 36.51 & +28 23 56.0 & 40.0 \\
    Inside10         & 0124710701 & 24 Jun 2000 & 12 57 27.68 & +28 08 41.0 & 27.2 \\
    Inside11         & 0124710301 & 27 Jun 2000 & 12 58 32.19 & +27 24 12.0 & 28.6 \\
    Inside12         & 0124712201 & 09 Dec 2000 & 12 57 42.51 & +27 43 38.0 & 27.6 \\
    Inside13         & 0124712001 & 10 Dec 2000 & 12 58 50.01 & +27 58 52.0 & 22.8 \\
    Inside14         & 0124712101 & 10 Dec 2000 & 12 57 27.68 & +28 08 41.0 & 28.1 \\
    Inside15         & 0124710801 & 10 Dec 2000 & 13 01 25.60 & +27 43 53.0 & 29.8 \\
    Inside16         & 0153750101 & 04 Dec 2001 & 12 59 46.71 & +27 56 60.0 & 25.8 \\
    Inside17         & 0124712401 & 05 Jun 2002 & 13 01 50.19 & +28 09 28.0 & 27.6 \\
    Inside18         & 0124712501 & 07 Jun 2002 & 13 00 36.50 & +28 25 15.0 & 28.7 \\
    Inside19         & 0204040101 & 06 Jun 2004 & 13 00 22.21 & +28 24 03.0 & 101.9 \\
    Inside20         & 0204040201 & 18 Jun 2004 & 13 00 22.21 & +28 24 03.0 & 108.3 \\
    Inside21         & 0204040301 & 12 Jul 2004 & 13 00 22.21 & +28 24 03.0 & 104.2 \\
    Inside22         & 0300530701 & 06 Jun 2005 & 12 59 36.92 & +27 58 14.8 & 25.5 \\
    Inside23         & 0300530601 & 07 Jun 2005 & 12 59 35.41 & +27 56 33.3 & 25.7 \\
    Inside24         & 0300530501 & 08 Jun 2005 & 12 59 39.73 & +27 55 12.0 & 25.5 \\
    Inside25         & 0300530401 & 09 Jun 2005 & 12 59 46.67 & +27 55 12.0 & 27.5 \\
    Inside26         & 0300530301 & 11 Jun 2005 & 12 59 51.00 & +27 56 33.3 & 31.0 \\
    Inside27         & 0300530201 & 17 Jun 2005 & 12 59 49.45 & +27 58 14.8 & 27.5 \\
    Inside28         & 0300530101 & 18 Jun 2005 & 12 59 43.18 & +27 58 59.8 & 25.5 \\
    Inside29         & 0304320301 & 27 Jun 2005 & 13 00 22.21 & +28 24 03.0 & 55.9 \\
    Inside30         & 0304320201 & 28 Jun 2005 & 13 00 22.21 & +28 24 03.0 & 80.8 \\
    Inside31         & 0304320801 & 06 Jun 2006 & 13 00 22.21 & +28 24 03.0 & 63.8 \\
    Inside32         & 0403150201 & 11 Jun 2006 & 12 57 42.51 & +27 19 09.7 & 55.2 \\
    Inside33         & 0403150101 & 14 Jun 2006 & 12 57 42.51 & +27 19 09.7 & 54.4 \\
    Inside34         & 0652310701 & 16 Jun 2010 & 12 57 24.29 & +27 29 52.0 & 21.8 \\
    Inside35         & 0652310201 & 18 Jun 2010 & 12 57 24.29 & +27 29 52.0 & 22.3 \\
    Inside36         & 0652310301 & 20 Jun 2010 & 12 57 24.29 & +27 29 52.0 & 19.4 \\
    Inside37         & 0652310401 & 24 Jun 2010 & 12 57 24.29 & +27 29 52.0 & 23.9 \\
    Inside38         & 0652310501 & 04 Jul 2010 & 12 57 24.29 & +27 29 52.0 & 22.9 \\
    Inside39         & 0652310601 & 06 Jul 2010 & 12 57 24.29 & +27 29 52.0 & 20.8 \\
    Inside40         & 0652310801 & 03 Dec 2010 & 12 57 24.29 & +27 29 52.0 & 16.9 \\
    Inside41         & 0652310901 & 05 Dec 2010 & 12 57 24.29 & +27 29 52.0 & 16.9 \\
    Inside42         & 0652311001 & 11 Dec 2010 & 12 57 24.29 & +27 29 52.0 & 16.4 \\
    Inside43         & 0691610201 & 02 Jun 2012 & 12 57 24.65 & +27 29 42.7 & 37.9 \\
    Inside44         & 0691610301 & 04 Jun 2012 & 12 57 24.65 & +27 29 42.7 & 35.9 \\
    Inside45         & 0851180501 & 30 May 2019 & 13 02 00.14 & +27 46 57.8 & 48.4 \\
    \hline
    \end{tabular}
    \label{tab: xmm_observations}
\end{table*}


\bsp	
\label{lastpage}
\end{document}